\pdfoutput=1
\documentclass[preprint]{aastex63}
\usepackage{amsmath,amssymb}

\usepackage{color}

\usepackage{dcolumn}
\newcolumntype{d}[1]{D{.}{.}{#1}}

\received{XXXX}
\revised{XXXX}
\accepted{XXXX}

\submitjournal{ApJS}

\shorttitle{MARVEL C{\rm a}OH}
\shortauthors{Wang et al.}

\newcommand{\Cv}[1]{${\mathcal C}_{#1{\rm v}}$}

\newcommand{\cm}{cm$^{-1}$}

\begin{document}
\title{MARVEL analysis of the measured high-resolution rovibronic spectra of the calcium monohydroxide radical (CaOH)}

\correspondingauthor{Sergei N. Yurchenko}
\email{s.yurchenko@ucl.ac.uk}

\author{Yixin Wang}
\affiliation{Department of Physics and Astronomy, University College London, Gower Street, WC1E 6BT London, UK; Nankai University, 94 Weijin Road, Tianjin, China}

\author[0000-0002-5167-983X]{Alec Owens}
\affiliation{Department of Physics and Astronomy, University College London, Gower Street, WC1E 6BT London, UK}


\author[0000-0002-4994-5238]{Jonathan Tennyson}
\affiliation{Department of Physics and Astronomy, University College London, Gower Street, WC1E 6BT London, UK}

\author[0000-0001-9286-9501]{Sergei N. Yurchenko}
\affiliation{Department of Physics and Astronomy, University College London, Gower Street, WC1E 6BT London, UK}

\begin{abstract}
The calcium monohydroxide radical (CaOH) is an important astrophysical molecule relevant to cool stars and rocky exoplanets, amongst other astronomical environments. Here, we present a consistent set of highly accurate rovibronic (rotation-vibration-electronic) energy levels for the five lowest electronic states ($\tilde{X}\,^2\Sigma^+$, $\tilde{A}\,^2\Pi$, $\tilde{B}\,^2\Sigma^+$, $\tilde{C}\,^2\Delta$, $\tilde{D}\,^2\Sigma^+$) of CaOH. A comprehensive analysis of the published spectroscopic literature on this system has allowed 1955 energy levels to be determined from 3204 rovibronic experimental transitions, all with unique quantum number labelling and measurement uncertainties. The dataset covers rotational excitation up to $J=62.5$ for molecular states below 29\,000~cm$^{-1}$. The analysis was performed using the MARVEL algorithm, which is a robust procedure based on the theory of spectroscopic networks. The dataset provided will significantly aid future interstellar, circumstellar and atmospheric detections of CaOH, as well as assisting in the design of efficient laser cooling schemes in ultracold molecule research and precision tests of fundamental physics.
\end{abstract}


\section{Introduction}
\label{sec:intro}

The calcium monohydroxide radical ($^{40}$Ca$^{16}$O$^{1}$H) is a linear triatomic molecule of increasing astronomical interest due to its expected presence in the atmospheres of hot rocky super-Earth exoplanets~\citep{09Bernath.exo,jt693}. This class of exoplanets are very close to their host star and tidally-locked, with their dayside exposed to extremely high temperatures, e.g.\ 2000--4000~K. The material present on the surface of the planet, including rock-forming elements such as silicon, magnesium, iron, calcium, and so on, will vaporise to some extent and produce an atmosphere strongly dependant on planetary composition~\citep{12ScLoFe.exo,16FeJaWi}. Investigating the spectroscopy of hot rocky super-Earths requires accurate spectroscopic data on simple molecules composed of rock-forming elements, like calcium monohydroxide. However, data for CaOH is not necessarily available or easily accessible. For example, a recent systematic study modelling M-dwarf photospheres by \citet{13RaReAl.CaOH} noted missing opacity from the benchmark BT-Settl model due to three molecules: NaH, AlH and CaOH. The ExoMol project has since computed line lists for NaH~\citep{jt605} and
AlH~\citep{jt732} meaning that CaOH, notably its band around 18\,000~cm$^{-1}$, remains as the only identified missing source of opacity in these objects.

Given the high cosmic abundance of calcium with respect to molecular hydrogen, it is reasonable to expect calcium-bearing molecules such as CaOH in other interstellar and circumstellar environments. For example, a possible formation mechanism in the interstellar medium (ISM) is through the reaction of Ca$^+$ ions with small oxide interstellar grains to release gas-phase CaOH~\citep{78DuMixx.CaOH}. \citet{73Tsuji.CaOH} predicted that CaOH would be the most abundant calcium-bearing molecule in oxygen-rich late-type stars at temperatures of $T=1000$--$2000$~K. While the $\tilde{B}\,^2\Sigma^+$--$\tilde{X}\,^2\Sigma^+$ electronic band of CaOH was tentatively assigned in the spectra of late-type M-dwarf stars~\citep{72Pesch.CaOH}.

A large number of experimental studies have measured the rovibronic (rotation-vibration-electronic) spectrum of CaOH, however, there is no centralised source containing this information, aside from the CDMS database~\citep{CDMS:2001,CDMS:2005,CDMS} but this only covers the microwave region (0--34~cm$^{-1}$). In this work, we present a dataset of highly accurate molecular rovibronic transitions and energy levels for the CaOH molecule, obtained by evaluating all available spectroscopic data on CaOH from the published literature using the MARVEL (Measured Active Rotational-Vibrational Energy Levels) algorithm~\citep{jt412,07CsCzFu.method,12FuCsxx.methods,jt750}. This procedure takes a set of assigned transition frequencies with measurement uncertainties and converts it into a consistent set of empirical energy levels, each with their own  measurement uncertainty and unique quantum numbers and state labels.

A MARVEL dataset for CaOH will considerably aid future astronomical detection of this molecule, particularly because of its large wavelength and rotational excitation coverage. Furthermore, it will benefit the calculation of a molecular line list for CaOH, which is currently being undertaken by the ExoMol project~\citep{jt528,jt631}. MARVEL datasets of empirical energy levels can greatly improve the accuracy of computed molecular opacities, as was done for the recent MARVEL titanium oxide dataset~\citep{jt672} and TiO line list~\citep{jt760}, whose detection in exoplanet atmospheres had been hampered by the inaccuracy of line positions in the available line lists~\citep{15HoDeSn.TiO}.

We mention that the alkaline earth monohydroxide radicals, including CaOH~\citep{19KoStYu.CaOH,19AuBoxx.CaOH}, are relevant in studies of ultracold molecules and precision tests of fundamental physics due to their favourable energy level structure. A list of highly accurate energy levels across multiple electronic states can be useful in this field, especially for the design of efficient laser cooling schemes which requires knowledge of molecular rovibronic structure to a high degree of accuracy.

\section{Theoretical background}
\label{sec:theory}

\subsection{The MARVEL approach}

The MARVEL procedure~\citep{jt412,07CsCzFu.method,12FuCsxx.methods,jt750} is based on the theory of spectroscopic networks~\citep{11CsFuxx.methods,12FuCsxx.methods,14FuArMe.methods,16ArFuCs.methods} and offers an elegant way to construct and represent complex networks such as those contained in a molecule's spectroscopy. Energy levels are represented as nodes with the allowed transitions linking them and the corresponding transition intensities acting as weights. Provided with a dataset of assigned transitions with measurement uncertainties, MARVEL will produce a consistent set of uniquely labelled empirical-quality energy levels, with the uncertainties propagated from the input transitions to the output energies.

From a user-perspective, transitions included in the MARVEL dataset must have a measurement uncertainty and every energy level has to be uniquely labelled, typically by a set of quantum numbers as discussed below. The chosen set of quantum numbers must be consistent across the whole dataset but need not be physically meaningful. However, a sensible choice will benefit comparisons with other data and allow the final dataset to be readily utilized in future studies. MARVEL is publicly available through a user-friendly web interface at \href{http://kkrk.chem.elte.hu/marvelonline}{http://kkrk.chem.elte.hu/marvelonline} and numerous MARVEL studies have been performed on astronomically important diatomic and small polyatomic molecules: NH$_3$~\citep{jt608}, C$_2$~\citep{jt637}, TiO~\citep{jt672}, C$_2$H$_2$~\citep{jt705}, H$_2$S~\citep{jt718}, ZrO~\citep{jt740}, NH~\citep{jt764}, SO$_2$~\citep{jt704}, H$_3^+$~\citep{13FuSzMa.H3+} and isotopologues~\citep{13FuSzFa.H3+}, and H$_2{}^{16}$O and its isotopologues~\citep{jt750,jt562,jt454,jt482,jt539,jt576}.

\subsection{Electronic structure and spectroscopy of CaOH}

\begin{figure}[ht]
\includegraphics[width=\textwidth]{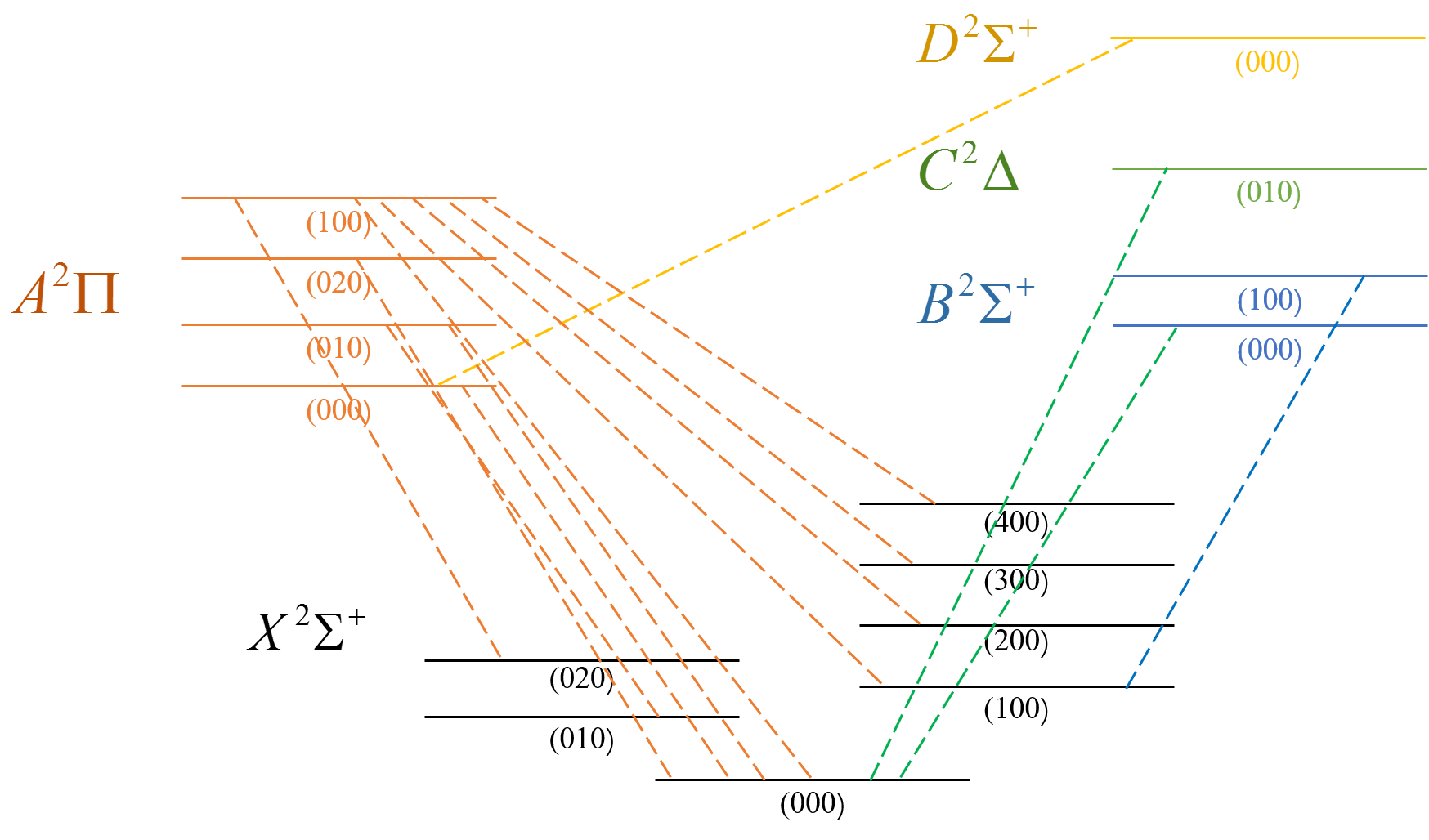}
\caption{The rovibronic states and transitions of CaOH considered in this work.}
\label{fig:bands}
\end{figure}

The calcium monohydroxide radical is an open-shell system with a relatively complex electronic structure. To date, only the lowest-lying eight electronic states up to the $\tilde{G}\,^2\Pi$ state~\citep{97HaJaBe.CaOH} at approximately 32\,633~cm$^{-1}$ are known. In this work we consider the lowest five electronic states ($\tilde{X}\,^2\Sigma^+$, $\tilde{A}\,^2\Pi$, $\tilde{B}\,^2\Sigma^+$, $\tilde{C}\,^2\Delta$, $\tilde{D}\,^2\Sigma^+$) and the transitions linking them, as shown in Figure~\ref{fig:bands}. This choice was governed by the availability of laboratory rovibronic transtion data in the literature. The wavenumber regions considered for each state were: 0--2599~cm$^{-1}$ ($\tilde{X}\,^2\Sigma^+$); 15\,966--17\,677~cm$^{-1}$ ($\tilde{A}\,^2\Pi$); 18\,023--18\,849~cm$^{-1}$ ($\tilde{B}\,^2\Sigma^+$); 22\,197--23\,457~cm$^{-1}$ ($\tilde{C}\,^2\Delta$); 28\,157--28\,898~cm$^{-1}$ ($\tilde{D}\,^2\Sigma^+$).

Interestingly, the spectrum of CaOH is affected by the Renner-Teller effect, see e.g.\ \citet{19Jungen} for a recent review, which is caused by the interaction of electronic orbital and vibrational angular momenta in linear molecules. This effect manifests itself when the molecule bends by lifting the degeneracy of the electronic states, for example, the first excited $\tilde{A}\,^2\Pi$ state of CaOH is split into two components $A^{\prime}$ and $A^{\prime\prime}$ at bent configurations. The Renner-Teller effect complicates spectral analysis due to the increased energy level congestion in band systems but its treatment is necessary for a correct description and several experimental studies have considered Renner splittings in CaOH~\citep{95LiCoxx.CaOH,92LiCoxx.CaOH,92JaBexx.CaOH,91CoLiPr.CaOH,83HiQiHa.CaOH}.

\subsection{Vibronic coupling, symmetry and quantum numbers}

MARVEL requires that each transition is between energy levels with unique state labels and quantum numbers that are consistent across the entire dataset. For CaOH, different authors have used different combinations of quantum numbers, as listed in Table~\ref{t:QN}. In this work we have selected eight quantum numbers, shown in Table~\ref{t:QN:MARVEL}, that form a consistent set and allow each state to be uniquely labelled. The electronic state and the vibronic state are labelled along with the rotational angular momentum quantum number $J$, the rotationless parity $e$/$f$,  and the quantum labels $F_1$ and $F_2$ denoting spin components $J = N + 1/2$ and $J = N - 1/2$, respectively. Normal mode notation $(v_1,v_2^{L},v_3)$ is used for the vibrational states, where $v_1, v_2, v_3$ represent the symmetric stretch, bending, and asymmetric stretch modes. The quantum number $L$ will be used to refer to the absolute value of the vibrational angular momentum quantum number $l$ associated with the $\nu_2$ bending mode, $L=|l|$, where the vibrational quantum number $l$ takes the following values:
$$
|l| = v_2, v_2-2, v_2-4, \ldots, 0\,({\rm or}\,1).
$$
For example, the $v_2 = 1$ state has two degenerate components $l = \pm 1$. The $v_2 = 2$ state assumes three states of $l = \pm 2$ and $l=0$. The $v_2 = 3$ state splits into four components $l = \pm 3$ and $l=\pm 1$ and so on. The typical designation of the linear bending quantum mode therefore includes $|l|$: $(v_1, v_2^{|l|},v_3)$. For $v_2 = 0$ and often for $v_2 = 1$ the superscript $|l|$ can be omitted. Coupling with other degrees of freedom (such as  molecular rotation) lifts the  degeneracy of the $l\ne 0 $ states.\footnote{Here we use the linear triatomic version of the bending quantum number $v_2$, which is related to the bent molecule case as  $v_2 \equiv  v_2^{\rm linear} = 2 v_{2}^{\rm bent} + L$,
where $L$ is constrained to $L = K_a  \pm \Lambda$ \citep{98BuJexx}, and $K_a = |k_a|$ is the rotational quantum number associated with the projection of the total rotational angular momentum on the $a$ axis (or $z$ axis in our case).}

\begin{table}
\caption{State labels and quantum numbers used by different authors for energy levels of CaOH.}
\label{t:QN}
\begin{tabular}{ll}
\hline\hline
Label & Description\\
\hline
Electronic state	& Electronic state, e.g.\ $\tilde{X}\,^2\Sigma^+$ \\
Vibronic state	& Vibronic state, e.g.\ $\tilde{X}\,^2 \Delta$, $\mu\, \tilde{A}\,^2 \Pi$, $\kappa\, \tilde{A}\,^2 \Pi$  \\
$J$	& Rotational angular momentum\\
$e/f$	& Total (rotationless Kronig) `parity' \\
$c/d$	& Mulliken's label\\
$v_1$	& Symmetric stretching mode \\
$v_2$	& Bending mode (linear molecules)\\
$l$	& Vibrational angular momentum associated with $\nu_2$ mode\\
$v_3$	&  Antisymmetric stretching mode\\
$\mu$ and $\kappa$ & Two vibronic components of $\tilde{A}\,^2 \Pi$ \\
$F_1/F_2$	& Spin components $J = N + 1/2$ or $J = N - 1/2$\\
$\Omega$ & Projection of the total angular momentum on the $z$ axis. \\
\hline\hline
\end{tabular}
\end{table}

\begin{table}
\caption{State labels and quantum numbers of CaOH adopted for the MARVEL dataset.}
\label{t:QN:MARVEL}
\begin{tabular}{ll}
\hline\hline
Label & Description\\
\hline
Vibronic state	& Vibronic state, e.g. muA2Sigma ($\mu\,\tilde{A}\,^2\Sigma^+$) or X2Pi ($\tilde{X}\,^2\Pi$)  \\
$J$	& Rotational angular momentum\\
$e/f$	& Rotationless parity \\
$v_1$	& Symmetric stretching mode \\
$v_2$	& Bending mode (linear molecules)\\
$L$	& Vibrational angular momentum associated with $\nu_2$ mode ($L=|l|$)\\
$v_3$	&  Antisymmetric stretching mode\\
$F_1/F_2$	& Spin components $J = N + 1/2$ or $J = N - 1/2$\\
\hline\hline
\end{tabular}
\end{table}

When the molecule is linear, the \Cv{\infty}(M) molecular symmetry group is used to classify the rotation, vibration and electronic degrees of freedom. The lowest five electronic states are $\tilde{X}\,^2\Sigma^+$, $\tilde{A}\,^2\Pi$, $\tilde{B}\,^2\Sigma^+$, $\tilde{C}\,^2\Delta$, $\tilde{D}\,^2\Sigma^+$ with the projections of the electronic angular momentum on the $z$ axis $\Lambda = 0$ ($\tilde{X}$, $\tilde{B}$, $\tilde{D}$), $\pm 1$ ($\tilde{A}$) and $\pm 2$ ($\tilde{C}$). The symmetry of the vibrational motion is controlled by the vibrational angular momentum quantum number $L$ in the same way that the symmetry of the electronic state is defined by $\Lambda$: $\Sigma$, $\Pi$, $\Delta$, \ldots, for $l =0,\pm 1,\pm 2,\ldots,$ respectively.
The electronic angular momentum can be coupled to the vibrational angular momentum giving the vibronic angular momentum with projections $\Lambda+l$, which are also described using the notation $\Sigma^{\pm}, \Pi, \Delta, \Phi, \ldots$. These vibronic symmetries are then used to label the final vibronic states
while the electronic symmetries are omitted. Consider, for instance, the vibrational state $(0,1^1,0)$  in the first excited electronic state
 $\tilde{A}\,^2\Pi$. Coupling its vibrational angular momentum with $l=\pm 1$ ($\Pi$) and the electronic angular momentum with $\Lambda = \pm 1$ leads to the three vibronic components $|\Lambda+l| = 0,0,2$, i.e. $\Sigma^{+}$, $\Sigma^{-}$ and $\Delta$, with a doubly degenerate state $\Delta$.  These vibronic states are then assigned $\tilde{A} (0,1^1,0)$ $^2\Sigma^{\pm}$ and  $\tilde{A}(0,1^1,0)$ $^2\Delta$.

The bending motion of the molecule ($v_2>0$) lifts the electronic degeneracy of $\Lambda\ne 0$ states ($\tilde{A}\,^2\Pi$ and $\tilde{C}\,^2\Delta$) via the Renner-Teller interaction ($\Lambda$ doubling), see, e.g., \citet{60Pople,62Hougen.method,62Hougen.methoda,62Hougen.methodb}.
There are two alternative notations for these components  used in the literature. Experimental studies of CaOH tend to use $\mu$ and $\kappa$ for the two vibronic sub-bands associated with the Renner-Teller splitting, while a more general consideration denotes the two Renner-Teller components $A'$ and $A''$ of the $C_s$ symmetry group to classify the molecular states, as well as the different degrees of freedom of CaOH. According to this convention, the $\kappa$ levels are always higher in energy than the $\mu$ levels of the same $J$ \citep{62Hougen.methodb}.
In the above example of $(0,1^1,0)$ in the excited electronic state $\tilde{A}\,^2\Pi$, the final vibronic notations inherit the  symmetry of a linear molecule and are given by $\tilde{A}(0,1^1,0)$  $\mu/\kappa$ $^2\Sigma^{\pm}$ and  $\tilde{A}(0,1^1,0)$  $^2\Delta$.

\subsection{Symmetry considerations}

When coupling different angular momenta of CaOH, it is common to first couple the electronic and vibrational angular momenta towards the vibronic momentum with projections $\Lambda+l$ (also assigned $^2\Sigma^{\pm}, \Pi, \Delta, \Phi, \ldots$) and then with the unpaired spin ($S=1/2$) towards the total angular momentum with the projection $\Omega = \Lambda+l+\Sigma$, where $\Sigma$ is the projection of the electron spin angular momentum
on the $z$ axis\footnote{Note that this label $\Sigma$ is different from the label $\Sigma$ used to assign the $\Lambda=0$ state.}. For CaOH, it is common to use the  $F_1$ and $F_2$ quantum labels to distinguish the two spin components with $|\Lambda+l|\pm 1/2$. For example, in the $^2\Sigma$ states the rotational angular momentum $\bf{N}$ is coupled to spin as $\bf{J} = \bf{N} + \bf{S} $ and thus their projections on the $z$ axis (at linear configurations) are related by $J=N\pm 1/2$, where $N$ is the rotational angular momentum. Each rotational state of $^2\Sigma$  consists of two spin sub-components, assigned $F_1$ and $F_2$.  $\Omega$ is sometimes used as an alternative to $F_1$ and $F_2$  to identify vibronic sub-components, e.g. as $^2\Pi_{1/2}$ and $^2\Pi_{3/2}$, $^2\Delta_{3/2}$ and $^2\Delta_{5/2}$  etc. Most of the sources presented in this work, however, use the $F_1$ and $F_2$ notation.

The symmetry of the rovibronic state is a product of the symmetries of the electronic, vibrational,  spin and rotational parts and can also be associated with the corresponding  angular momenta $\bf{G}$,  $\bf{L}$, $\bf{S}$ and $\bf{N}$. However, the final rovibronic state can only be  $\Sigma^+$ (parity is $+$) or $\Sigma^-$ (parity is $-$) of the molecular symmetry group \Cv{\infty}(M)~\citep{70Watson.method}.
The parity of  $^2\Sigma^+$ or $^2\Sigma^-$ states are $(-1)^{N}$ and $(-1)^{N+1}$, respectively. The state parity always changes for the electric dipole transitions $+ \leftrightarrow -$, which can be used to reconstruct the parity of the upper/lower state if only one of them is known. The rotationless $e/f$ parity is related to the $+/-$ parity as follows: levels with parity  $+(-1)^{J-0.5} $ are $e$ levels, and levels with parity  $-(-1)^{J-0.5} $ are $f$ levels~\citep{75BrHoHu.diatom}. Some authors use the Mulliken' labels $c, d$ \citep{31Mulliken} to distinguish states of different parities, for example in the case of $v_2=3$~\citep{96ZiFlAn.CaOH} with $(0,3^{1c},0)$, $(0,3^{1d},0)$, $(0,3^{3c},0)$ and $(0,3^{3d},0)$.

The electric dipole selection rules are
\begin{align}
J' &= J'' \pm 1;  \\
  e &\leftrightarrow  e \quad {\rm and} \quad f \leftrightarrow  f \quad  {\Delta J = \pm 1}; \\
  e &\leftrightarrow  f \quad \Delta J = 0; \\
  \Sigma^+ &\leftrightarrow \Sigma^- \quad (\rm total~symmetry).
\end{align}

\subsection{Examples of coupling vibration, spin and electronic angular momenta}

For non-$\Sigma$ states, the interaction with the electron spin can involve the vibrational and electronic angular momentum.
For example, the vibrational excitation $(0,1^1,0)$ of the electronic state $\tilde{X}\,^2\Sigma^+$ gives rise to the total vibronic angular momentum of the symmetry $^2\Pi$ ($l=\pm 1$), referenced to as $\tilde{X}(0,1^1,0)\,^2\Pi$. The resulting rovibronic energy pattern forms a quartet, which can be designated using the four combinations of $e,f$ and $F_1,F_2$.

Including the electronic angular momentum complicates this picture even more. For example, in order to couple all angular momenta in the vibronic state $\tilde{A}(0,1^1,0)\,^2\Pi$, in this work we first couple the electronic and vibrational momenta with projections $l=\pm 1$ and $\Lambda=\pm 1$, respectively, towards the vibronic angular momentum with $l+\Lambda = 0^{+/-}, \pm 2$. The zero vibronic components $0^{+/-}$ have two symmetries $^2\Sigma^+$ and $^2\Sigma^-$ with $|\Omega| = 1/2$. In this case $F_1$ and  $F_2$ are redundant and can be chosen to match $e$ and $f$ states. The state with $l+\Lambda = \pm 2$ has the symmetry $^2\Delta$ with  $|\Omega|$ taking values $3/2$ and $5/2$, and therefore can also be designated $(0,1^1,0){}^2\Delta_{3/2}$ ($F_1$) and $(0,1^1,0){}^2\Delta_{5/2}$ ($F_2$), respectively, where the labels $F_1$ and $F_2$ are assigned to the $|\Omega| = 1/2$ and $3/2$ states based on the order of the corresponding energy values.

Consider another example of the $\tilde{A}(0,2,0)\,^2\Pi$ state ($\Lambda=\pm 2$, $l=0,\pm 2$) with the vibronic angular momenta $\Lambda+l = \pm 1$ and $\pm 3$, which represent the vibronic states $^2\Pi$ and $^2\Phi$, respectively. The $\Lambda = \pm 2$ degeneracies are lifted due to the interaction with the bending  mode (Renner-Teller) leading to the $\mu$ and $\kappa$ sub-components (or $A'$ and $A''$). The interaction with  spin introduces the spin splitting with the final values of the projections of the angular momenta $1/2$ and $3/2$ ($^2\Pi$) or $5/2$ and $7/2$ ($^2\Phi$), i.e. to $^2\Pi_{1/2}$, $^2\Pi_{3/2}$, $^2\Phi_{5/2}$ and $^2\Phi_{7/2}$. However, if $l=0$ but $\Lambda\ne 0$,  such as e.g. for the $\tilde{A}(0,0,0)\,^2\Pi$ state, the final projections of the total angular momentum $\Omega = \Lambda + \Sigma$ are $\pm 1/2$ and $\pm 3/2$, and these states can be assigned $^2\Pi_{1/2}$ and $^2\Pi_{3/2}$, respectively.




\section{Experimental data sources}

Spectroscopic data was extracted from thirteen published sources~\citep{06DiShWa.CaOH,96ZiFlAn.CaOH,93ScFlSt.CaOH,95LiCoxx.CaOH,94CoLiPr.CaOH,92LiCoxx.CaOH,92CoLiPr.CaOH,92ZiBaAn.CaOH,92JaBexx.CaOH,91CoLiPr.CaOH,85BeBrxx.CaOH,84BeKixx.CaOH,83HiQiHa.CaOH}. These data are summarised in Table~\ref{tab:sources}. Only data from \citet{06DiShWa.CaOH} was provided in digital format while all other literature sources had to be processed using digitisation software. A unique reference label was assigned to each extracted transition, which is a requirement for the MARVEL input file. The reference indicates the data source, Table (or page) and line number that the transition originates from. The data source tag, for example 06DiShWa, is based on the notation employed by the IUPAC task group on water~\citep{jt482,jt562}.

Aside from the numerous experimental measurements of CaOH spectra,  there have been several studies investigating the electronic structure of CaOH using quantum chemical methods~\citep{84BaPaxx.CaOH,86BaLaPa.CaOH,90BaLaSt.CaOH,90Ortiz.CaOH,96KoBoxx.CaOH,02KoPexx.CaOH,02ThPeLi.CaOH,05TaChFr.CaOH} but these have largely focused on molecular structures and properties rather than  rovibronic spectroscopy.

\begin{table}[t]
\label{tab:sources}
\caption{Experimental sources of CaOH spectra and their coverage.}
\centering
\resizebox{\textwidth}{!}{%
\footnotesize
\begin{tabular}{lllllll}
\hline\hline
\multicolumn{1}{l}{Experimental Source. } & \multicolumn{1}{c}{Electron. States} & \multicolumn{1}{c}{$(v_1\,v_2\,v_3)$\footnote{Vibrational quantum numbers for upper and lower states (see text)}} & \multicolumn{1}{c}{$J$ range} & \multicolumn{1}{c}{Coverage (cm$^{-1}$)}	&	\multicolumn{1}{c}{A/V\footnote{Available/Verified number of lines from literature source}} & \multicolumn{1}{c}{$\Delta$ (cm$^{-1}$)\footnote{Average uncertainty of transition wavenumber data}} \\
	\hline

	83HiQiHa~\citep{83HiQiHa.CaOH} & $\tilde{A}\,^2\Pi$--$\tilde{X}\,^2\Sigma^+$ & (000)--(000) & 1.5--47.5 & 15,928--16,045 & 131/115 & 0.3\\

	84BeKi & $\tilde{B}\,^2\Sigma^+$--$\tilde{X}\,^2\Sigma^+$ & (000)--(000) & 1.5--26.5 & 18,008--18,036  & 73/73 & 0.01\\
\citep{84BeKixx.CaOH}	 & $\tilde{B}\,^2\Sigma^+$--$\tilde{X}\,^2\Sigma^+$ & (100)--(100) & 5.5--25.5 & 18,007--18,033  & 48/48 & 0.01\\

	85BeBr~\citep{85BeBrxx.CaOH} & $\tilde{A}\,^2\Pi$--$\tilde{X}\,^2\Sigma^+$ & (000)--(000) & 1.5--53.5 & 15,970--16,068 & 155/155 & 0.01\\

	91CoLiPr~\citep{91CoLiPr.CaOH} & $\tilde{A}\,^2\Pi$--$\tilde{X}\,^2\Sigma^+$ & (100)--(000) & 0.5--56.5 & 16,578--16,659  & 201/201 & 0.005\\

	92LiCo~\citep{92LiCoxx.CaOH} & $\tilde{A}\,^2\Pi$--$\tilde{X}\,^2\Sigma^+$ & (020)--(000) & 1.5--40.5 & 16,641--16,807 & 322/322 & 0.01\\
	
	92CoLiPr~\citep{92CoLiPr.CaOH} & $\tilde{A}\,^2\Pi$--$\tilde{X}\,^2\Sigma^+$ & (100)--$(02^00)$ & 4.5--29.5  & 15,951--15,988  & 25/25 & 0.104\\
     & $\tilde{A}\,^2\Pi$--$\tilde{X}\,^2\Sigma^+$ & (100)--$(02^20)$ & 4.5--26.5  & 15,909--15,960  & 48/48 & 0.056\\
	 & $\tilde{A}\,^2\Pi$--$\tilde{X}\,^2\Sigma^+$ & (100)--(100) & 5.5--29.5 & 15,957--16,069 & 54/54 & 0.064 \\
	 & $\tilde{A}\,^2\Pi$--$\tilde{X}\,^2\Sigma^+$ & (100)--(200) & 4.5--29.5 & 15,352-- 15,392 & 42/42 & 0.056 \\
	 & $\tilde{A}\,^2\Pi$--$\tilde{X}\,^2\Sigma^+$ & (100)--(300) & 4.5--27.5 & 14,762--14,798 & 30/30 & 0.17 \\
	 & $\tilde{A}\,^2\Pi$--$\tilde{X}\,^2\Sigma^+$ & (100)--(400) & 14.5--25.5 & 14,206--14,214 & 9/9 & 0.10\\
	
	92ZiBaAn~\citep{92ZiBaAn.CaOH} & $\tilde{X}\,^2\Sigma^+$--$\tilde{X}\,^2\Sigma^+$ & (000)--(000) & 2.5--16.5 & 2.67--10.69 & 44/44 & 1.7$\times 10^{-6}$ \\
	
	92JaBe~\citep{92JaBexx.CaOH} & $\tilde{X}\,^2\Sigma^+$--$\tilde{X}\,^2\Sigma^+$ & (000)--(000) & 2.5--38.5 & 0.045--10.69 & 25/25 & 1$\times 10^{-5}$ \\
	 & $\tilde{C}\,^2\Delta$--$\tilde{X}\,^2\Sigma^+$ & (010)--(000) & 0.5--62.5 & 0.045--22,247 & 373/374 & 0.33\\

	94CoLiPr~\citep{94CoLiPr.CaOH} & $\tilde{A}\,^2\Pi$--$\tilde{X}\,^2\Sigma^+$ & $(01^10)$--$(01^10)$ & 1.5--38.5 & 15,923--16,095 & 409/409 & 0.06 \\	
	 & $\tilde{A}\,^2\Pi$--$\tilde{X}\,^2\Sigma^+$ & $(01^10)$--(000) & 0.5--41.5 & 16,289--16,461 & 409/409 & 0.016 \\	
	
	93ScFlSt~\citep{93ScFlSt.CaOH} & $\tilde{X}\,^2\Sigma^+$--$\tilde{X}\,^2\Sigma^+$ & (000)--(000) & 0.5--3.5 & 0.67--2.01 & 26/26 & 3.3$\times 10^{-7}$\\

	95LiCo~\citep{95LiCoxx.CaOH} & $\tilde{A}\,^2\Pi$--$\tilde{X}\,^2\Sigma^+$ & $(01^10)$--$(01^10)$ & 1.5-61.5 & 15,935--16,428 & 640/640 & 0.01\\	
	 & $\tilde{A}\,^2\Pi$--$\tilde{X}\,^2\Sigma^+$ & $(01^10)$--(000) & 40.5--60.5 & 15,935--16,428 & 640/640 & 0.01\\

	96ZiFlAn~\citep{96ZiFlAn.CaOH} & $\tilde{X}\,^2\Sigma^+$--$\tilde{X}\,^2\Sigma^+$ & (000)--(000) & 2.5--19.5 & 2.67--12.69 & 28/28 & 5$\times 10^{-6}$\\
	 & $\tilde{X}\,^2\Sigma^+$--$\tilde{X}\,^2\Sigma^+$ & $(01^10)$--$(01^10)$ & 12.5--19.5 & 9.32--12.64 & 23/23 & 5$\times 10^{-6}$\\
	 & $\tilde{X}\,^2\Sigma^+$--$\tilde{X}\,^2\Sigma^+$ & $(02^00)$--$(02^00)$ & 12.5--19.5 & 9.31--12.64 & 36/36 & 8.3$\times 10^{-6}$\\
	 & $\tilde{X}\,^2\Sigma^+$--$\tilde{X}\,^2\Sigma^+$ & $(02^20)$--$(02^20)$ & 12.5--19.5 & 9.31--12.64 & 36/36 & 8.3$\times 10^{-6}$\\
     & $\tilde{X}\,^2\Sigma^+$--$\tilde{X}\,^2\Sigma^+$ & $(100)$--$(100)$ & 12.5--19.5 & 9.29--12.61 & 12/12 & 8.3$\times 10^{-6}$ \\

	 06DiShWa~\citep{06DiShWa.CaOH}& $\tilde{A}\,^2\Pi$--$\tilde{X}\,^2\Sigma^+$ & (000)--(000) & 1.5--53.5 & 15,928--16,068 & 267/267 & 0.017\\	
	 & $\tilde{D}\,^2\Sigma^+$--$\tilde{A}\,^2\Pi$ & (000)--(000) & 0.5--45.5 & 12,108--12,232 & 206/206 & 0.018\\

	\hline\hline
\end{tabular}}
\end{table}

\subsection{Comments on literature sources}

\textbf{83HiQiHa}: \citet{83HiQiHa.CaOH} contains transitions from the $\tilde{A}(0,0,0)\,^2\Pi$--$\tilde{X}(0,0,0)\,^2\Sigma^+$ band. Most of these data are included in the more recent study by \citet{06DiShWa.CaOH} but some of the line positions show deviations up to 0.3~cm$^{-1}$. For example, the transition $\tilde{A}(0,0,0)\,^2\Pi$, $J=36.5$ $F=2$, $e$ $\leftarrow$ $\tilde{X}(0,0,0)\,^2\Pi$, $J=36.5$ $F=2$, $f$ appears in \citet{83HiQiHa.CaOH} as 16\,029.986 \cm\ and in \citet{06DiShWa.CaOH} as 16\,030.285~\cm.  We believe that the recent data~\citep{06DiShWa.CaOH} is more reliable.  In these instances we have removed the lines by \citet{83HiQiHa.CaOH} from our MARVEL analysis by changing the sign of the corresponding frequency value (MARVEL convention).

\textbf{84BeKi}: \citet{84BeKixx.CaOH} contains transitions from the $\tilde{B}(0,0,0)\,^2\Sigma^+$--$\tilde{X}(0,0,0)\,^2\Sigma^+$ and $\tilde{B}(1,0,0)\,^2\Sigma^+$--$\tilde{X}(1,0,0)\,^2\Sigma^+$ band.

\textbf{85BeBr}: \citet{85BeBrxx.CaOH} contains transitions from the $\tilde{A}(0,0,0)\,^2\Pi$--$\tilde{X}(0,0,0)\,^2\Sigma^+$ band.

\textbf{91CoLiPr}: \citet{91CoLiPr.CaOH} contains transitions from the $\tilde{A}(1,0,0)\,^2\Pi$--$\tilde{X}(0,0,0)\,^2\Sigma^+$ band.

\textbf{92LiCo}: \citet{92LiCoxx.CaOH} reported  transitions from the $\tilde{A}(0,2^0,0)\,^2\Pi$--$\tilde{X}(0,0,0)\,^2\Sigma^+$ band system.

\textbf{92CoLiPr}: \citet{92CoLiPr.CaOH} studied the $\tilde{A}\,^2\Pi$--$\tilde{X}\,^2\Sigma^+$ system, covering the $\tilde{A}(1,0,0)\,^2\Pi$--$\tilde{X}(1,0,0)\,^2\Sigma^{+}$, $\tilde{A}(1,0,0)\,^2\Pi$--$\tilde{X}(2,0,0)\,^2\Sigma^{+}$, $\tilde{A}(1,0,0)\,^2\Pi$--$\tilde{X}(3,0,0)\,^2\Sigma^{+}$, $\tilde{A}(1,0,0)\,^2\Pi$--$\tilde{X}(4,0,0)\,^2\Sigma^{+}$, $\tilde{A}(1,0,0)\,^2\Pi$--$\tilde{X}(0,2^0,0)\,^2\Sigma^+$ and $\tilde{A}(1,0,0)\,^2\Pi$--$\tilde{X}(0,2^2,0)\,^2\Delta$ vibronic bands.

\textbf{92ZiBaAn}: \citet{92ZiBaAn.CaOH} investigated eleven pure rotational transitions in the $\tilde{X}(0,0,0)\,^2\Sigma^+$ ground state.  The total angular momentum quantum number $F$ used to account for hyperfine structure was not included in our dataset. Instead, the uncertainty of these transitions was increased to match the hyperfine splittings. The dataset was converted to cm$^{-1}$ from the original units of MHz.

\textbf{92JaBe}: \citet{92JaBexx.CaOH} studied the $\tilde{C}\,^2\Delta$--$\tilde{X}\,^2\Sigma^+$ system covering the $\tilde{C}(0,1^1,0)\,^2\Pi$--$\tilde{X}(0,0,0)\,^2\Sigma^+$ band.


\textbf{93ScFlSt}: \citet{93ScFlSt.CaOH} investigated the hyperfine structure of the three lowest pure rotational transitions of the $\tilde{X}(0,0,0)\,^2\Sigma^+$ state. The total angular momentum quantum number $F$ used to account for hyperfine structure was not included in our dataset.  Instead, the uncertainties for these transitions were increased to 0.0001~\cm\ to account for the fact that we have neglected hyperfine effects.




\textbf{94CoLiPr}: \citet{94CoLiPr.CaOH} measured transitions in the $\tilde{A}\,^2\Pi$--$\tilde{X}\,^2\Sigma^+$ band system:  $\tilde{A}(0,1^1,0)\,^2\Sigma^+$--$\tilde{X}(0,1^1,0)\,^2\Pi$ and $\tilde{A}(0,1^1,0)\,^2\Sigma^-$--$\tilde{X}(0,0,0)\,^2\Sigma^+$ bands.
The upper doubling states are assigned $\kappa\,^2\Sigma^-$ or $\mu\,^2\Sigma^+$.

\textbf{95LiCo}: \citet{95LiCoxx.CaOH} studied the $v_2 = 1$ bending vibrational levels of the $\tilde{A}\,^2\Pi$ and $\tilde{X}\,^2\Sigma^+$ states, covering transitions of the following bands: $\tilde{A}(0,1^1,0)\kappa\,^2\Sigma^-$, $^2\Delta, \mu\,^2\Sigma^+$ $\leftarrow$ $\tilde{X}(0,1^1,0)\,^2\Pi$, $(0,0,0)\,^2\Sigma^+$.
It should be noted that $\tilde{A}(0,1^1,0)$--$\tilde{X}(0,0,0)$ is electric dipole forbidden. Even though the $\tilde{A}(0,1,0)\mu{}^2\Sigma^+$ and $\tilde{A}(0,1^1,0)\kappa{}^2\Sigma^-$ states can be correlated to $A'$ and $A''$, we retained the experimental labels $\mu$ and $\kappa$. The $e/f$ parity of all states was reconstructed from the $+/-$ parities $p$, which in turn were obtained from the parity $p''$ of the lower vibronic state $\Sigma^+$ as $(-1)^{N}$. For non-$\Sigma$  lower vibronic states ($\tilde{X}\,^2\Pi$) we used the upper state parity instead if the vibronic symmetries were either $\Sigma^+$ ($\mu$) or $\Sigma^-$ ($\kappa$): the parities $p'$ are $(-1)^{N'}$ or $-(-1)^{N'}$, respectively.

\textbf{96ZiFlAn}: \citet{96ZiFlAn.CaOH} has pure rotational and rovibrational transitions in the $\tilde{X}\,^2\Sigma^+$ ground electronic state: $(0,0,0)$, $(0,1^{1c},0)$, $(0,1^{1d},0)$, $(0,2^{0},0)$, $(0,2^{2c},0)$, and $(0,2^{2d},0)$ where the superscript $c/d$ represents $l$-type doubling effects. The $(0,1^{1},0)$  and $(0,2^{2},0)$ vibronic states of $\tilde{X}\,^2\Sigma^+$ have the vibrational symmetries $^2\Pi$ and $^2\Delta$, respectively.  The correlation between $c/d$ and $e/f$ does not appear to follow the rules given by \citet{75BrHoHu.diatom}. In order to reconstruct this correlation we used the  $(0,1^{1},0)$ spectroscopic constants from \citet{96ZiFlAn.CaOH} in the program PGOPHER~\citep{PGOPHER}, assuming the vibronic symmetry $^2\Pi$, and computed the corresponding transitions allowing us to correlate the $e/f$ and $c/d$ parities as follows:
for $c$,  $F_1=e$ and $F_2=f$; for $d$, $F_1=f$ and $F_2=e$, {but only for the $(0,0,0)$, $(0,1^{1c},0)$, $(0,1^{1d},0)$ states. The CaOH spectroscopic constants of $\tilde{X}(0,2^{2},0)\,^2\Delta$ by \citet{96ZiFlAn.CaOH} only work with PGOPHER if assuming the symmetry $^2\Pi$ instead of $^2\Delta$, which  in this case worked with the same conversion rules as for the $(0,1^1,0)$ transitions. Apparently this is due to the effective  rotational Hamiltonian model used in the fit in the original work. The dataset was converted to cm$^{-1}$ from the original units of MHz.

\textbf{06DiShWa}: \citet{06DiShWa.CaOH} covers transitions from the $\tilde{A}(0,0,0)\,^2\Pi$--$\ \tilde{X}(0,0,0)\,^2\Sigma^+$ and $\tilde{D}(0,0,0)\,^2\Sigma^+$--$\tilde{A}(0,0,0)\,^2\Pi$ bands. In their analysis, \citet{06DiShWa.CaOH} used the $\tilde{X}(0,0,0)\,^2\Sigma^+$--$\tilde{X}(0,0,0)\,^2\Sigma^+$ transitions from the work by \citet{92ZiBaAn.CaOH}. A number of the $\tilde{A}$--$\tilde{X}$ transitions were measured in the work of \citet{83HiQiHa.CaOH}.

\section{Results and Discussion}

Our final MARVEL transition file (input, see Table~\ref{tab:trans}) consists of 3204 experimental (excluding the 16 transitions from \citet{83HiQiHa.CaOH} deemed less reliable) and 20 pseudo-experimental (PGOPHER) transitions and has the following structure:
\begin{center}
\begin{tabular}{llrrrrrrrrrl}
$\tilde{\nu}$& unc &QN$'$ & QN$''$ & source$_i$ \\
\end{tabular}\\
\end{center}
where $\tilde{\nu}$ is the transition wavenumber (\cm), unc is the experimental uncertainty (\cm),  QN$'$ and QN$''$ are the quantum numbers of the upper and lower states, respectively, and `source' is the abbreviation of the literature source concatenated with a counting number $i$ of the data from this source. source$_i$ is a unique ID of the transitions in the MARVEL dataset. The pseudo-experimental transition wavenumbers were reconstructed with the PGOPHER program using the spectroscopic constants from \citet{95LiCoxx.CaOH} and \citet{96ZiFlAn.CaOH} to support the corresponding datasets compensating for missing lines. Their uncertainties were set to 1~\cm\ in order not to interfere with the true experimental values.  The QN set comprises 8 quantum numbers/labels selected to represent a general rovibronic state of CaOH as in Tables~\ref{tab:trans} and \ref{tab:energies}: Vibronic state label, $e/f$ parity, $J$, $v_1$, $v_2$, $L$, $v_3$, and  $F = F_1, F_2$.

The MARVEL energy file (output) was processed via the online MARVEL app using the Cholesky (analytic) approach with a 0.05~cm$^{-1}$ threshold on the uncertainty of the ``very bad'' lines, which produced a MARVEL energy file containing 1955 states. The MARVEL dataset covers rotational excitation up to $J=62.5$ for molecular states below 29\,000~cm$^{-1}$. The MARVEL output and structure of the MARVEL energy levels is illustrated in Table~\ref{tab:energies} and also plotted in Fig~\ref{fig:en} as a function of $J$.

\begin{figure}
	\includegraphics[width=1.0\linewidth]{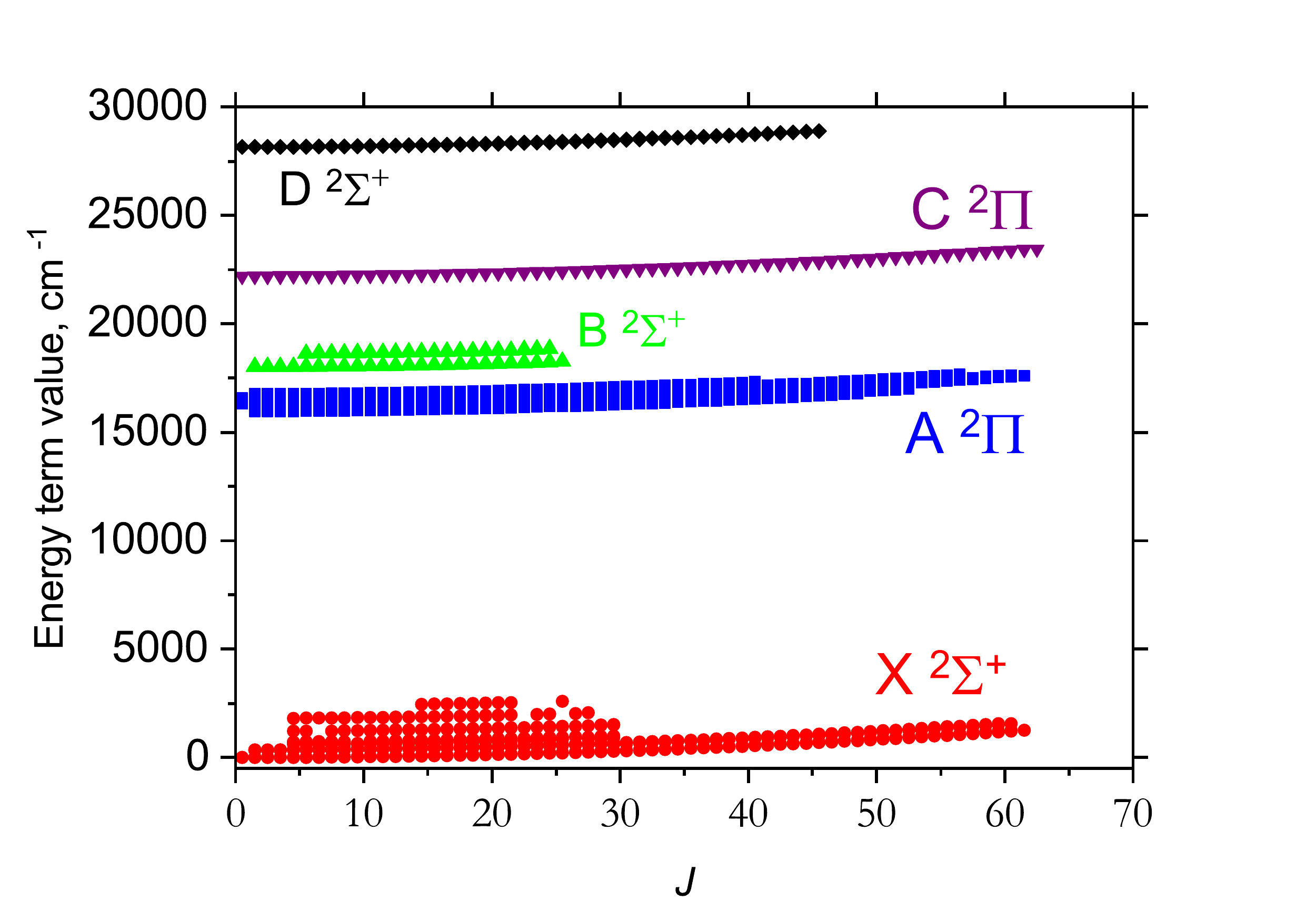}
	\caption{The MARVEL energy term values for CaOH shown for different electronic bands.}
	\label{fig:en}
\end{figure}

\begin{table}[t]
\label{tab:trans}
\caption{Extract from the MARVEL transition file. The quantum numbers/labels are described in Table~\protect\ref{t:QN:MARVEL}.
The MARVEL frequency wavenumber $\tilde{\nu}$ and uncertainties are in \cm.}
\centering
{\tt
\scriptsize
\tabcolsep=3pt
\begin{tabular}{d{3}d{3}crcrrrrc crcrrrrcl}
\hline\hline
\multicolumn{1}{c}{ $\tilde{\nu}$} & \multicolumn{1}{c}{\rm unc.,\cm} &  \multicolumn{8}{c}{\rm Quantum `numbers' of upper states}  &
                                                 \multicolumn{8}{c}{\rm Quantum `numbers' of lower states}  & {\rm Source }   \\
 & & {\rm Vibronic}$'$  & $J'$ &$e'/f'$& $v'_1$ & $v'_2$ & $L'$ &$v'_3$ & $F'_1/F'_2$ &
     {\rm Vibronic}$''$  & $J''$ &$e''/f''$& $v''_1$ & $v''_2$ & $L''$ &$v''_3$ & $F''_1/F''_2$ &  \\
	\hline
     16026.295 &   0.005 &   A2Pi  &     3.5 &f &   0 & 0 & 0 & 0&  F2  &      X2Sigma+ &    4.5& f  &  0 & 0 & 0 & 0 & F2 &  06DiShWa1     \\
     16025.364 &    0.01 &   A2Pi  &     4.5 &f &   0 & 0 & 0 & 0&  F2  &      X2Sigma+ &    5.5& f  &  0 & 0 & 0 & 0 & F2 &  06DiShWa2     \\
     16024.464 &   0.005 &   A2Pi  &     5.5 &f &   0 & 0 & 0 & 0&  F2  &      X2Sigma+ &    6.5& f  &  0 & 0 & 0 & 0 & F2 &  06DiShWa3     \\
     16023.571 &   0.005 &   A2Pi  &     6.5 &f &   0 & 0 & 0 & 0&  F2  &      X2Sigma+ &    7.5& f  &  0 & 0 & 0 & 0 & F2 &  06DiShWa4     \\
     16022.696 &    0.01 &   A2Pi  &     7.5 &f &   0 & 0 & 0 & 0&  F2  &      X2Sigma+ &    8.5& f  &  0 & 0 & 0 & 0 & F2 &  06DiShWa5     \\
     16021.842 &   0.005 &   A2Pi  &     8.5 &f &   0 & 0 & 0 & 0&  F2  &      X2Sigma+ &    9.5& f  &  0 & 0 & 0 & 0 & F2 &  06DiShWa6     \\
     16021.002 &   0.005 &   A2Pi  &     9.5 &f &   0 & 0 & 0 & 0&  F2  &      X2Sigma+ &   10.5& f  &  0 & 0 & 0 & 0 & F2 &  06DiShWa7     \\
      16020.18 &   0.005 &   A2Pi  &    10.5 &f &   0 & 0 & 0 & 0&  F2  &      X2Sigma+ &   11.5& f  &  0 & 0 & 0 & 0 & F2 &  06DiShWa8     \\
     16019.375 &   0.005 &   A2Pi  &    11.5 &f &   0 & 0 & 0 & 0&  F2  &      X2Sigma+ &   12.5& f  &  0 & 0 & 0 & 0 & F2 &  06DiShWa9     \\
     16018.586 &   0.005 &   A2Pi  &    12.5 &f &   0 & 0 & 0 & 0&  F2  &      X2Sigma+ &   13.5& f  &  0 & 0 & 0 & 0 & F2 &  06DiShWa10    \\
     16017.815 &   0.005 &   A2Pi  &    13.5 &f &   0 & 0 & 0 & 0&  F2  &      X2Sigma+ &   14.5& f  &  0 & 0 & 0 & 0 & F2 &  06DiShWa11    \\
\hline
\hline
\end{tabular}
}
\end{table}

\begin{table}[t]
\label{tab:energies}
\caption{Extract from the MARVEL energy file. The quantum numbers/labels are described in Table~\protect\ref{t:QN:MARVEL}, which are followed by the MARVEL energy term value (\cm), uncertainty (\cm) and the number of transitions supporting the state in question.}
\centering
{\tt
\begin{tabular}{lrccccccd{6}d{6}c}
\hline\hline
\multicolumn{8}{c}{\rm Quantum `numbers'}  &\multicolumn{1}{c}{ $\tilde{E}$} & \multicolumn{1}{c}{\rm unc.} & {\rm no. of}   \\
Vibronic  & $J$ &$e/f$& $v_1$ & $v_2$ & $L$ &$v_3$ & $F_1/F_2$ &\multicolumn{1}{c}{\rm \cm}  & \multicolumn{1}{c}{\rm \cm} & {\rm trans.}  \\
	\hline
X2Delta      &   24.5  &  e     &   0   &  2   &  2  &   0  &  F1    &      911.045055    &       0.05    &   1  \\
X2Delta      &   24.5  &  f     &   0   &  2   &  2  &   0  &  F2    &      927.860974    &       0.05    &   1  \\
X2Delta      &   25.5  &  f     &   0   &  2   &  2  &   0  &  F2    &      944.898055    &       0.05    &   1  \\
X2Delta      &   26.5  &  f     &   0   &  2   &  2  &   0  &  F2    &      963.000974    &       0.05    &   1  \\
muA2Sigma    &    0.5  &  e     &   0   &  1   &  1  &   0  &  F1    &     16310.22824    &      0.005    &   1  \\
muA2Sigma    &    0.5  &  f     &   0   &  1   &  1  &   0  &  F2    &     16310.70825    &      0.005    &   1  \\
muA2Sigma    &    1.5  &  e     &   0   &  1   &  1  &   0  &  F1    &     16311.01157    &   0.007071    &   2  \\
muA2Sigma    &    1.5  &  f     &   0   &  1   &  1  &   0  &  F2    &     16311.96407    &   0.070711    &   2  \\
muA2Sigma    &    2.5  &  e     &   0   &  1   &  1  &   0  &  F1    &     16312.46245    &   0.003462    &   6  \\
muA2Sigma    &    2.5  &  f     &   0   &  1   &  1  &   0  &  F2    &       16313.911    &   0.003333    &   3  \\
muA2Sigma    &    3.5  &  e     &   0   &  1   &  1  &   0  &  F1    &      16314.6022    &   0.004994    &   2  \\
	\hline\hline
\end{tabular}
}
\end{table}

Figure~\ref{fig:CaOH_CDs} offers a visual representation of the CaOH MARVEL network, where the upper state energies are connected with the lower state energies via circles. The size of a circle is $\log(n+1)$, where $n$ is the number of transitions supporting the corresponding upper state.
The vertical bars along the horizontal-axis show the lower state energies, while the horizontal bars along the vertical-axis give the upper state energies. The value of $n$ ranges from 1 (dark blue) to 38 (red).

\begin{figure}
	\includegraphics[width=1.0\linewidth]{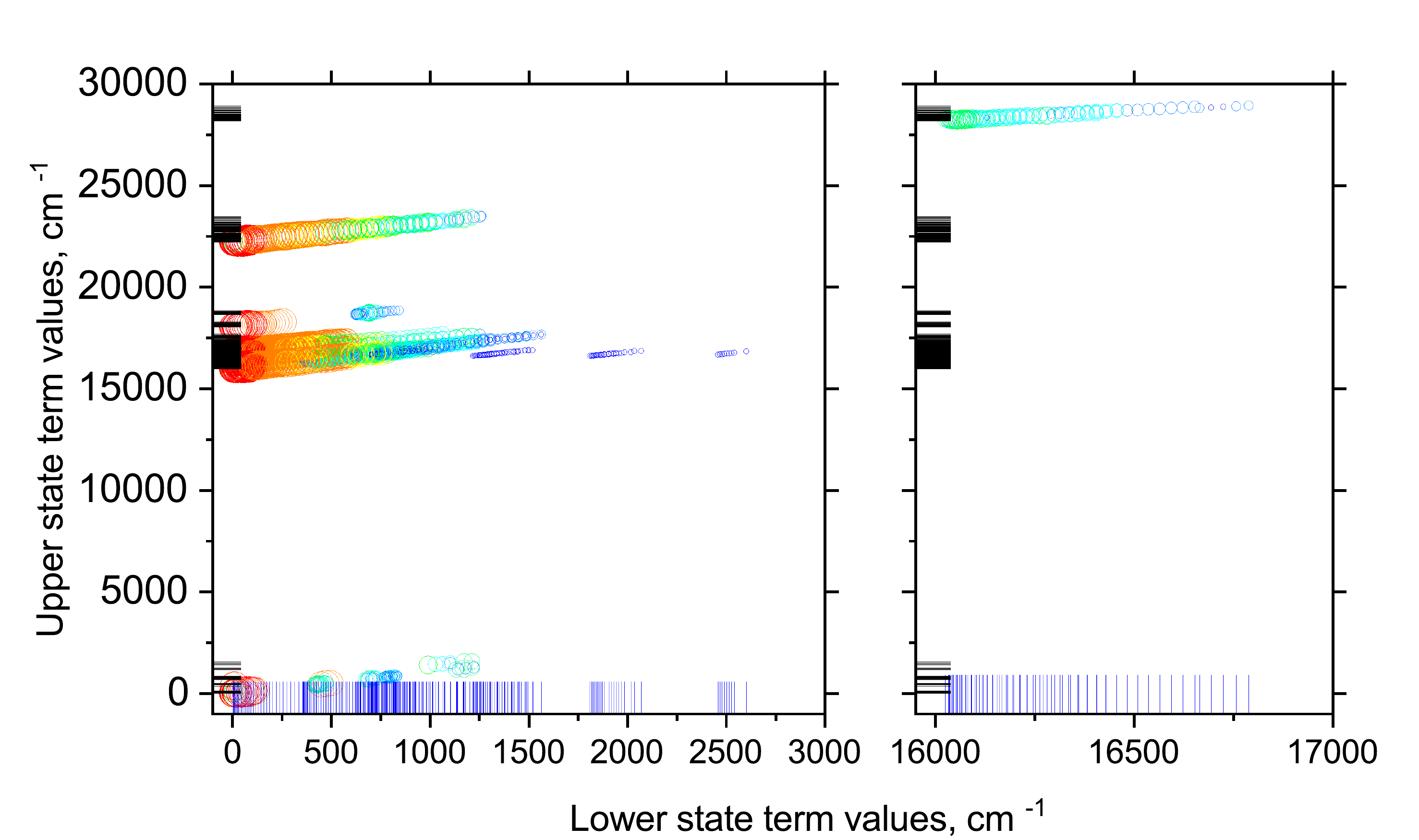}
	\caption{The MARVEL network for CaOH: the upper state energies are plotted against corresponding lower state energies. The vertical bars along the horizontal-axis represent the lower state energies, while the horizontal bars along the vertical-axis give the upper state energies. Each circle represents a particular transition, with the size proportional to the log of the number of transitions $n$ plus 1, going to the upper state. The value ranges from 1 (dark blue) to 38 (red).}
	\label{fig:CaOH_CDs}
\end{figure}

\section{Conclusions}
\label{sec:conc}

We have comprehensively evaluated the published spectroscopic literature on CaOH and extracted all meaningful molecular rovibronic transition data. These data were analysed using the robust MARVEL algorithm which converts assigned transitions into a consistent set of uniquely labelled empirical energy levels with measurement uncertainties. The dataset covers rotational excitation up to $J=62.5$ for 1955 molecular states below 29\,000~cm$^{-1}$. The MARVEL input and output files for CaOH are provided as supplementary material and can be readily updated to include new experimental rovibronic measurements.

While we have analyzed data from five electronic states and a large range of rotational levels, the experimental data has only limited coverage of vibrationally excited states. There is no empirical information on the $\nu_3$ stretching vibrational mode and only limited information on vibrational excitation of the other modes in electronically excited states. This means that any line list constructed for this molecule will have to rely on {\it ab initio} predictions for the missing quantities.

With the renewed interest in CaOH we expect the new MARVEL dataset to help future detection of this molecule in astronomical environments. The most immediate benefit will be in the calculation of a comprehensive CaOH molecular line list as part of the ExoMol project~\citep{jt528,jt631}. A list of highly accurate empirical energy levels is necessary to refine the theoretical spectroscopic model of a molecule to achieve orders-of-magnitude improvements in the accuracy of the predicted line positions. Accurate molecular opacities of CaOH are essential for detecting this molecule in hot rocky super-Earth exoplanets and this work brings us closer to this goal.

As mentioned before, a detailed knowledge of the energy level structure in CaOH will help the design of efficient laser cooling schemes in ultracold molecule research and precision tests of fundamental physics. The alkaline earth monohydroxide radicals are attractive molecules in this pursuit due to their favourable energy level structure, as demonstrated by SrOH, which was the first untrapped polyatomic molecule to be laser-cooled~\citep{17KoBaMa.SrOH}.

\acknowledgments%
This work was supported by the STFC Project No. ST/R000476/1.
We thank Peter Bernath for collecting experimental data and many helpful discussions.
Yixin Wang's visit was supported by Physics Boling Class in Nankai University.

\software{MARVEL, PGOPHER}

\bibliographystyle{aasjournal}
\bibliography{journals_astro,jtj,exoplanets,CaOH,linelists,CH4,H3+,SrOH,Renner,programs,methods,books,diatomic,TiO}

\end{document}